\documentclass[12pt]{article}
\usepackage{graphicx}
\usepackage{bm}
\def\be{\begin{equation}}
\def\ee{\end{equation}}
\def\bea{\begin{eqnarray}}
\def\eea{\end{eqnarray}}
\usepackage{latexsym}
\usepackage{epsfig,amssymb,euscript}
\usepackage{amsmath}
\usepackage{array,calc,epsfig}
\newcommand{\Tr}{{\rm Tr}}

\renewcommand\( {\left(}
\renewcommand\) {\right)}

\newcommand\nn{\nonumber\\}

\def\beq{\begin{eqnarray}}
\def\eeq{\end{eqnarray}}

\def\bi{\begin{itemize}}
\def\ei{\end{itemize}}

\def\ex#1{\,{\rm e}^{#1}}
\def\inv{^{\raise.15ex\hbox{${\scriptscriptstyle -}$}\kern-.05em 1}}
\newcommand{\sst}[1]{\scriptscriptstyle{#1}}

\renewcommand{\d}{\delta}

\renewcommand{\r}{\rho}

\newcommand{\s}{\sigma}

\newcommand{\m}{\mu}
\newcommand{\n}{\nu}

\newcommand{\ka}{\kappa}
\renewcommand{\l}{\lambda}

\newcommand{\om}{\omega}

\newcommand\cO{\mathcal O}

\newcommand{\rh}{r_{\kern-.2em\sst{H}}}
\newcommand{\rhoh}{\r_{\raise-.25ex\hbox{$\kern-.1em\sst{H}$}}}

\newcommand\del{\partial}

\newcommand\rmd{{\rm d}}
\newcommand\rmi{{\,\rm i\,}}

\newcommand{\dsq}{\rmd s^2}

\begin{document}
\baselineskip=15.5pt
\pagestyle{plain}
\setcounter{page}{1}

\newfont{\namefont}{cmr10}
\newfont{\addfont}{cmti7 scaled 1440}
\newfont{\boldmathfont}{cmbx10}
\newfont{\headfontb}{cmbx10 scaled 1728}
\renewcommand{\theequation}{{\rm\thesection.\arabic{equation}}}

\vspace{1cm}

\begin{center}
{\huge{\bf On the Universality of the\\ Chern-Simons Diffusion Rate}}
\end{center}

\vskip 10pt

\begin{center}
{\large Francesco Bigazzi$^{a}$, Aldo L. Cotrone$^{a,b}$, Flavio Porri$^{a,b}$}
\end{center}

\vskip 10pt
\begin{center}
\textit{$^a$ INFN, Sezione di Firenze; Via G. Sansone 1, I-50019 Sesto Fiorentino
(Firenze), Italy.}\\
\textit{$^b$ Dipartimento di Fisica e Astronomia, Universit\`a di
Firenze; Via G. Sansone 1, I-50019 Sesto Fiorentino
(Firenze), Italy.}\\

\vspace{0.2cm}
{\small bigazzi@fi.infn.it, cotrone@fi.infn.it, porri.flavio@gmail.com}
\end{center}

\vspace{25pt}

\begin{center}
 \textbf{Abstract}
\end{center}

\noindent 

We prove the universality of the Chern-Simons diffusion rate - a crucial observable for the chiral magnetic effect - in a large class of planar strongly correlated gauge theories with dual string description. 
When the effects of anomalies are suppressed, the diffusion rate is simply given in terms of temperature, entropy density and gauge coupling, with a universal numerical coefficient. 
We show that this result holds, in fact, for all the top-down holographic models where the calculation has been performed in the past, even in presence of magnetic fields and anisotropy. 
We also extend the check to further well known models for which the same computation was lacking. 
Finally we point out some subtleties related to the definition of the Chern-Simons diffusion rate in the presence of anomalies. In this case, the usual definition of the rate - a late time limit of the imaginary part of the retarded correlator of the topological charge density - would give an exactly vanishing result, due to its relation with a non-conserved charge correlator. We confirm this observation by explicit holographic computations on generic isotropic black hole backgrounds. Nevertheless, a non-trivial Chern-Simons relaxation time can in principle be extracted from a quasi-normal mode calculation. 
\newpage

\section{Introduction}
\setcounter{equation}{0}

The Chern-Simons diffusion rate $\Gamma_{CS}$ is a fundamental observable in the study of chiral effects in the Quark-Gluon Plasma (QGP).
Its value, setting the rate of (local) change of the Chern-Simons number, determines, through the axial anomaly, the (local) change rate of chirality imbalance in the plasma.
Chirality imbalance is then responsible, in presence of a magnetic field, for the chiral magnetic effect \cite{Fukushima:2008xe}, which can be in principle measured in current experiments.

Being inherently non perturbative, the Chern-Simons diffusion rate cannot be computed in QCD from first principles. 
In the literature there are interesting effective field theory calculations, setting for example the behavior with $N$ (the number of colors of the theory) to be ${\cal O}(N^0)$ \cite{Moore:2010jd}.
But it is fair to say that we lack a solid estimate (analogous to what can be obtained from the lattice for other observables) for the value of $\Gamma_{CS}$ in the realistic strong coupling regime of the QGP. 

In this paper we address the problem of the calculation of the Chern-Simons diffusion rate by means of holographic techniques, as first performed in \cite{Son:2002sd}.
Thus, we consider the planar, strong coupling regime of quantum field theories which can model (but always present some differences from) real world QCD.
Holography has proven to give fruitful indications about the physics of transport coefficients in QCD.
In particular, when some universality can be found among different holographic models, the result for the observable under scrutiny can be used as a benchmark for the experiments, since clearly it does not depend on many of the details of the theory.
The shear viscosity is the most notable example of such a situation \cite{Policastro:2001yc,Kovtun:2004de}.

We are going to prove that at leading order in the planar strong coupling limit, the Chern-Simons diffusion rate is universal in the phenomenologically most relevant class of top-down holographic models.
This class of models includes, among the others, ${\cal N}=4$ SYM \cite{Son:2002sd}, even in presence of anisotropies \cite{matranca, bu} and magnetic fields \cite{Basar:2012gh}, the Witten-Sakai-Sugimoto model \cite{Witten:1998zw,SS,ben} and the Maldacena-Nu\~nez model \cite{Chamseddine:1997nm,mn}, for which the computation of $\Gamma_{CS}$ was lacking.  
In all the cases the field theory comes either from D3-branes or wrapped D$p$-branes with $p>3$.

More specifically, in absence of axial anomalies, the Chern-Simons diffusion rate is given, with a universal coefficient, in terms of the temperature $T$, entropy density $s$ and the gauge coupling $\alpha_s=g_{YM}^2/(4\pi)$ of the field theory
\be\label{result0}
\frac{\Gamma_{CS}}{s\, T} = \frac{\alpha_{s}^2(T)}{2^3\pi^3}\,.
\ee
At present there are no counterexamples to this formula in top-down holographic models. 
So, it is not excluded that its validity goes beyond the class of models considered in this paper.
Thus, the holographic behavior of this observable can be used with a certain degree of confidence as a benchmark for the QGP. 
For example, a very rough estimate of the critical temperature values $\alpha_{s}(T_c) \sim 1/2,\ s(T_c) \sim 10 T_c^3$ (lattice results without magnetic field \cite{Borsanyi:2013bia,Bazavov:2014pvz}) gives the value $\Gamma_{CS}/T_c^4 \sim 0.01$.
This would have quite a small effect in the QGP.

From a technical point of view, the derivation of the result (\ref{result0}) relies on the fact that the quadratic part of the five dimensional bulk action for the field dual to the topological charge density operator has a universal form in this class of models.
It is the action of a massless scalar.
Its kinetic term depends on a function which, in the models at hand, is the dual of the coupling constant (equation (\ref{acC})).
The universality of this result can be useful beyond the determination of the Chern-Simons diffusion rate.

The most interesting correction to the holographic result for the Chern-Simons diffusion rate in the planar expansion comes from possible anomalies proportional to the topological charge density operator.
In the bulk, these are described by a Stueckelberg action for the scalar mentioned above and a vector field dual to the anomalous current \cite{Klebanov:2002gr}. 
We are going to show that, in this case, the usual holographic prescription for the calculation of $\Gamma_{CS}$ gives an identically vanishing result.\footnote{All the results concerning $\Gamma_{CS}$ present in the literature boil down to the leading planar contribution, i.e. the effects of anomaly are not fully included in the calculation \cite{Drwenski:2015sha,Iatrakis:2015fma}.} 
This fact is very general for holographic field theories and does not rely on the details of the dual gravity backgrounds. 

As we are going to discuss, this feature is not related to the holographic limit, but it simply relies on the fact that the usual definition of $\Gamma_{CS}$, as the late time behavior of the (imaginary part of the) retarded correlator of the topological charge density, gives zero if the latter is proportional, as it happens due to the anomaly relation, to the retarded correlator of a non conserved (axial) charge. 
The relaxation time of the latter, can, instead, be consistently defined and expressed in terms of  the Chern-Simons diffusion rate computed with a cutoff in time i.e. effectively turning off the anomaly. 
Some considerations along this line can be already found in \cite{Moore:2010jd,Iatrakis:2015fma}. 
Chiral effects in presence of anomalies can be investigated, from a holographic point of view, through the analysis of quasi-normal modes of the dual black hole backgrounds. These quasi-normal modes, in fact, allow to extract the above mentioned relaxation time.

Before concluding this section, let us define the relevant quantities for our purposes.
The Chern-Simons diffusion rate is the probability of fluctuation of the Chern-Simons number $\Delta N_{CS}$ per unit time $t$ and unit volume $V$
\bea\label{def}
\Gamma_{CS} =  \frac{\langle (\Delta N_{CS})^2 \rangle}{V t} = \int d^4x \langle Q(x) Q(0) \rangle\,,
\eea
where the variation in the Chern-Simons number is
\begin{equation}
 \Delta N_{CS} = \int d^4x\,Q(x) = \int d^4x\,\frac{1}{16 \pi^2} {\rm Tr} F{\tilde F}\,.
\end{equation}
Here $F$ is the Yang-Mills field strength, $Q$ the topological charge density operator
\be
Q =\frac{1}{16 \pi^2} {\rm Tr} F{\tilde F}\,,
\ee
and we normalize the $SU(N)$ generators so that $\Tr [t^a\,t^b]=\delta^{a\,b}/2$ for the fundamental representation.

The two-point function in formula (\ref{def}) is the symmetrized Wightman one - everything is defined in real time. 
In a state at thermal equilibrium at temperature $T$, $\Gamma_{CS}$ can be given by a Kubo formula
\begin{equation}\label{CSdef}
\Gamma_{CS} = - \lim_{\omega \rightarrow 0} \frac{2 T}{\omega}  {\rm Im} G_R(\omega, \vec k =0)\,.
\end{equation}
Thus its computation boils down to the determination of the small frequency, zero momentum retarded correlator $G_R(\omega, \vec k =0)$ of the operator $Q$.
The retarded correlator is precisely what can be directly computed in holography from the standard prescriptions \cite{Son:2002sd}, once the bulk field dual to the operator $Q$ is identified.

The paper is organized as follows.
In section \ref{holoCS} we prove the universal formulae (\ref{result0}),  (\ref{acC}) for the Chern-Simons diffusion rate and the gravity action of the field dual to the corresponding operator in the phenomenologically relevant class of top-down holographic models in absence of anomalies.
Then, in section \ref{secanomaly} we discuss the effects of anomalies confirming, through a holographic computation on generic isotropic black hole backgrounds, that the standard prescription for the calculation of $\Gamma_{CS}$ gives an identically vanishing result. In turn, we discuss how to extract the Chern-Simons relaxation time from the quasi-normal modes of the dual black hole backgrounds.
The appendices include the analysis of the direct calculation of $\Gamma_{CS}$ in specific examples, including, for the first time, the Maldacena-Nu\~nez model. Moreover they contain a sketch of the derivation of a linear response formula for the axial relaxation time as well as specific AdS examples of the generic holographic results discusses in section \ref{secanomaly}.

\section{Holographic Chern-Simons diffusion rate}\label{holoCS}
\setcounter{equation}{0}

We are going to show that in the plasma phase of a large class of planar strongly correlated gauge theories with dual string description, the Chern-Simons diffusion rate reads
\be\label{result}
\frac{\Gamma_{CS}}{s\, T} = \frac{g_{YM}^4(T)}{2^7\pi^5}\,,
\ee
where the Yang-Mills coupling $g_{YM}$ and the entropy density $s$ are evaluated at the temperature scale $T$ of the plasma.
This result holds as long as any possible anomaly is ignored - effects of anomalies are discussed in the following section.
In the derivation below, we are going to prove another general result for the same class of models: the quadratic part of the five dimensional Lagrangian for the field dual to the topological charge density operator $Q$ is universal, being the one of a massless scalar times a function of the background metric which is precisely the (squared) coupling in the dual field theory.

\subsection{Derivation of the result}

The main difficulty in deriving a general result for the Chern-Simons (CS) diffusion rate in holography is that it requires a precise identification of the gravity fields dual to the topological charge density operator $Q$ and the Yang-Mills coupling $g_{YM}$.
Unlike what happens for the shear viscosity, whose related gravity field is a component of the metric which has a universal Lagrangian, the form of the five dimensional action for the gravity field dual to $Q$ is in principle model dependent.

Nevertheless, there exists a class of models where the identification of the field theory quantities needed for the computation is quite solid, allowing for the derivation of the result (\ref{result}).
The models we refer to are built up by wrapping D$p$-branes over $(p-3)$-cycles, giving rise at low energies to four dimensional gauge theories.
The case $p=3$, i.e. ${\cal N}=4$ SYM, is included in the discussion as well.
In these cases, we are going to show that the five dimensional action has indeed a universal form.

All the calculations of the Chern-Simons diffusion rate in holographic top-down models in the literature, and other ones we are going to discuss, are performed in representatives of this class of models.
As such, at present formula (\ref{result}) has no counter-example.
It is tempting to conjecture that it is valid for every theory with a gravity dual in the strong coupling regime.
As stated above, the difficulty in checking this statement in other models is due to the uncertain identification of the coupling and the field dual to $Q$.
It would be obviously interesting to provide a more general proof of the result (\ref{result}).

Let us now prove  (\ref{result}).
The main observation is that, in holographic models coming from D$p$-branes wrapped on $(p-3)$-cycles $\Omega_{p-3}$, the reduction of the brane DBI+WZ action
\be
S = -\tau_p \Tr \int d^{p+1}x\, e^{-\phi} \sqrt{-\det(g + 2\pi \alpha' F)} + \tau_p \Tr \int \sum_n C_n \wedge e^{2\pi\alpha' F}\,,
\ee
includes the Yang-Mills action
\be\label{ymac}
{S}_{YM} = -\int d^4 x \left[ \frac{1}{2 g_{YM}^2} \Tr F_{\mu\nu} F^{\mu\nu} + \frac{\theta_{YM}}{16 \pi^2} \Tr F_{\mu\nu} {\tilde F}^{\mu\nu} \right]\,,
\ee
with 
\bea\label{identifications}
&&\frac{1}{g_{YM}^2} =\frac{\tau_p}{2} (2\pi \alpha')^2 \int_{\Omega_{p-3}} d^{p-3}x\, e^{-\phi} \sqrt{\det g} \,, \nonumber \\
&&\theta_{YM} = (2\pi)^2\tau_p (2\pi \alpha')^2 \int_{\Omega_{p-3}} C_{p-3}\,.
\eea
In these formulae $g$ is the (pull-back of the) metric, $F$ the field strength of the gauge field on the brane world-volume and $C_n$ are RR $n$-form potentials; $\tau_p=(2\pi)^{-p} \alpha'^{-(p+1)/2}g_s^{-1}$ is the brane tension. 
The discussion is limited to models with vanishing NSNS $B$ field along the cycle $\Omega_{p-3}$. Notice that in the case $p=3$ (unwrapped D3-branes) the above relations give the usual identifications $g_{YM}^2 = 4\pi g_s$ and $\theta=2\pi C_0$, where $C_0$ is the type IIB axion.

We are going to consider Einstein frame actions in five dimension, so we rewrite the coupling as
\be\label{identificationE}
\frac{1}{g_{YM}^2} = \frac{\tau_p}{2} (2\pi \alpha')^2 \int_{\Omega_{p-3}} d^{p-3}x\, e^{\frac{p-7}{4}\phi} \sqrt{\det g_E} \,,
\ee
where $g_E$ is the Einstein frame metric.
Of course on shell the coupling does not depend on the frame.
In the probe brane spirit, the identifications above are going to be considered as the definitions of the couplings in the full theory where the branes have been replaced by geometry.

Inspection of formulae (\ref{ymac}) and (\ref{identifications}) dictates that the coupling of $Q = \frac{1}{16 \pi^2} {\rm Tr} F {\tilde F}$ with the dual gravity field $C$ is precisely $\int Q C$ if
\be\label{realC}
C=\tau_p (2\pi)^2 (2\pi \alpha')^2 \int_{\Omega_{p-3}} C_{p-3} \equiv  \tau_p (2\pi)^2 (2\pi \alpha')^2 {\rm Vol}(\Omega_{p-3})  {\tilde C}\,,
\ee
where ${\rm Vol}(\Omega_{p-3})$ is the volume of the cycle and we have introduced the reduced field ${\tilde C}$ which is the scalar typically present in the five dimensional reduced gravity action.
Thus, the gravity field dual to $Q$ is the $C_{p-3}$ RR field integrated on the $(p-3)$-cycle.

Let us derive its five dimensional action.
Consider the Einstein frame ten dimensional action for the corresponding field strength
\be\label{10d}
\frac{1}{2\kappa_{10}^2} \int d^{10}x \sqrt{-g_{10}}\, e^{\frac{7-p}{2}\phi} \left[ -\frac12  F_{p-2}^2 \right] \,,
\ee
together with the reduction ansatz
\be\label{redansatz}
ds^2_{10} = e^{f} ds^2_5 + ds^2_{\rm{int}}\,,
\ee
where $f$ is a function which encodes the dependence of the internal volume on the external coordinates. The reduced Einstein action has canonical form if 
\be
\int d^5y \sqrt{g_{\rm int}} = V_{\rm{int}} e^{-3f/2}\,,
\ee
where $y^i$, $i=1,\dots5$ are the internal coordinates and $V_{\rm{int}}$ is the constant part of the compactification volume.

Now, one of the indices of the $(p-2)$-form must be in the five-dimensional directions, while the other ones are along the $\Omega_{p-3}$ cycle.
Thus, assuming that the metric functions have a trivial dependence on the internal directions, we have
\be\label{effep}
F_{p-2}^2 = \partial_{M} {\tilde C} \partial^{M} {\tilde C} [\det(g_{E,\Omega'_{p-3}})]^{-1} e^{-f}\,.
\ee
With $\det(g_{E,\Omega'_{p-3}})$ we denote the determinant of the Einstein frame metric along the $\Omega_{p-3}$ cycle modulo its volume form, i.e. $\int_{\Omega_{p-3}} \sqrt{g_{E,\Omega_{p-3}}}= {\rm Vol}(\Omega_{p-3})\sqrt{\det(g_{E,\Omega'_{p-3}})}$.
``$M$'' is a five dimensional index.

In formula (\ref{effep}) we have assumed that there is no mixing of the field $\partial_{M}\tilde C$ with any vector potential: this amounts to assuming that anomalies are subleading effects, as it will become clear in the next sections.

Since
\be
\frac{1}{2\kappa_{10}^2}\int d^{10}x\sqrt{-g_{10}} = \frac{1}{2\kappa_{5}^2}\int d^5x \sqrt{-g_{5}}\, e^{f}\,, 
\ee
where $\frac{1}{2\kappa_{5}^2}=\frac{V_{\rm{int}}}{2\kappa_{10}^2}$, the reduction of (\ref{10d}) gives
\be\label{actilde}
\frac{1}{2\kappa_{5}^2} \int d^{5}x \sqrt{-g_{5}} \left[ -\frac12 \partial_{M} {\tilde C} \partial^{M} {\tilde C} \right] \left(\frac{1}{e^{\frac{p-7}{2}\phi} \det(g_{E,\Omega_{p-3}'})} \right)\,,
\ee
or equivalently 
\bea\label{acC}
&&\frac{1}{2\kappa_{5}^2} \int d^{5}x \sqrt{-g_{5}} \left[ -\frac12 \partial_{M} {\tilde C} \partial^{M} {\tilde C} \right] \left(\frac{{\rm Vol}(\Omega_{p-3})}{\int_{\Omega_{p-3}} e^{\frac{p-7}{4}\phi} \sqrt{\det g_E}} \right)^2= \\
&&\frac{1}{2\kappa_{5}^2} \int d^{5}x \sqrt{-g_{5}} \left[ -\frac12 \partial_{M} {C} \partial^{M} {C} \right] \frac{1}{(2\pi)^4}\left(\frac{1}{\tau_p (2\pi\alpha')^2  \int_{\Omega_{p-3}} e^{\frac{p-7}{4}\phi} \sqrt{\det g_E}} \right)^2 \,. \nonumber
\eea

From (\ref{identificationE}) one can recognize that the term in round parenthesis is nothing else than the fourth power of the coupling (times a number), so that we get the action
\be
\label{generalac}
\frac{1}{2\kappa_{5}^2} \int d^{5}x \sqrt{-g_{5}}\, H \left[ -\frac12 \partial_{M} C \partial^{M} C \right]\,,\qquad H= \left(\frac{g_{YM}^2}{8\pi^2}\right)^2 \,.
\ee
Thus, in the wrapped brane models, the quadratic part of the five dimensional action for the field dual to $Q$ is in a universal form. This is the main result of this section.

Now, for an action of the form (\ref{generalac}) - on a diagonal black hole metric with components depending only on the radial coordinate - with $H$ a function of the background fields and $C$ the field dual to the operator $Q$, the holographic calculation of the CS diffusion rate (following the original prescription in \cite{Son:2002sd}) gives \cite{kiritsis}
\be\label{general}
\Gamma_{CS} = \frac{1}{2\pi} H_{h}\, s\, T\,,
\ee
where in $H_h$ the fields are evaluated at the horizon and $s$ is the Bekenstein-Hawking entropy density. Since in our case (\ref{acC}) we have $H_h=\frac{g_{YM}^4(T)}{ (8\pi^2)^2}$, where the coupling is at the temperature scale $T$, (\ref{general}) gives immediately the result (\ref{result}).

Let us conclude this section by noting that the form of the action (\ref{actilde}) reproduces all the cases for which the calculation of the CS diffusion rate has been performed.
For ${\cal N}=4$ SYM it is immediate: there is no reduction and the dilaton is trivial,\footnote{In our conventions $g_s$ is included in $\tau_p$.} so $e^{\phi}=1=\det(g_{\Omega_{p-3}'})$ and we have the usual minimally coupled scalar studied in the original paper \cite{Son:2002sd}.
As it is shown in appendix \ref{appenkhar}, the same is true in the presence of a magnetic field \cite{Basar:2012gh}, whose only effect is to change the explicit form of the entropy density $s$ but not the form of the relation (\ref{result}). 
The same relation is also precisely satisfied by the Chern-Simons diffusion rate of the anisotropic ${\cal N}=4$ plasma of \cite{matranca} as computed in \cite{bu}. This is a relevant non trivial example since the background solution has a running dilaton. Details are provided in appendix \ref{appenmat}. 

We review the case of the wrapped D4-brane Witten's Yang-Mills (WYM) model \cite{Witten:1998zw} in appendix \ref{appendixwym} and derive (for the first time) the result for the ${\cal N}=1$ Super Yang-Mills model by Maldacena and Nu\~nez (MN) \cite{Chamseddine:1997nm,mn}, coming from wrapped D5-branes, in appendix \ref{appendixmn}. 
In all these cases the coupling, and so the ratio $e^{2\phi}/\det(g_{\Omega_{p-3}'})$ is a constant in the deconfined phase.
Finally, in the flavored version of the ${\cal N}=4$ SYM plasma \cite{Bigazzi:2009bk,Faedo:2016cih}, $\det(g_{\Omega_{p-3}'})=1$ and the dilaton factor in (\ref{acC}) is precisely what is given by the consistent reduction of the model \cite{Cotrone:2012um}, so $H_h \sim e^{2\phi_h} \sim g_{YM}^4(T)$.

It is worth noticing that the result (\ref{result}) holds at leading order in the holographic limits $N,\lambda\gg1$, where $\lambda=g_{YM}^2 N$ is the 't Hooft coupling. 
When the first higher derivative ($\alpha'$) corrections\footnote{See \cite{tranca} for holographic computations of the Chern-Simons diffusion rate in a Gauss-Bonnet setup.} are added on the gravity side - which amounts to include $1/\lambda^{3/2}$ corrections in the dual QFT -  the Chern-Simons diffusion rate of the ${\cal N}=4$ plasma is given by \cite{bu}
\be
\Gamma_{CS}= \frac{g_{YM}^4(T)}{2^7\pi^5}\frac{\pi^2}{2}N^2 T^4 \left[1- \frac{45}{8} \frac{\zeta(3)}{\lambda^{3/2}}+\cdots\right]\,,
\ee
while the entropy density is (see \cite{gkt})
\be
s = \frac{\pi^2}{2}N^2 T^3 \left[1+ \frac{15}{8} \frac{\zeta(3)}{\lambda^{3/2}}+\cdots\right]\,.
\ee
The result is that $\alpha'$ corrections reduce the Chern-Simons diffusion rate with respect to the value in (\ref{result}). This could led to conjecture that eq. (\ref{result}) sets an upper bound for $\Gamma_{CS}/s T$ for generic gauge theories.

A last comment is in order. 
We have restricted the discussion to models with no NSNS $B$ field turned on. 
A non trivial $B$ field is usually associated to multiple gauge groups, with more that one topological charge operator.
The prototype models are the Klebanov-Witten one \cite{Klebanov:1998hh} and its non conformal extension, the Klebanov-Strassler one \cite{Klebanov:2000hb}, which have two gauge groups.
It is immediate to verify that the field dual to the sum of the two topological charge operators has the same universal Lagrangian as above.

\section{Chern-Simons diffusion with $U(1)_A$ anomalies}\label{secanomaly}
\setcounter{equation}{0}
In a theory like QCD with $N_f$ massless quarks, the topological charge density operator $Q$ enters the anomaly equation for the axial current
\be
\partial_{\mu} J_A^{\mu} = - q Q\,,
\ee
where $q\sim N_f$ is the anomaly coefficient. 
This equation implies that the axial charge $Q_A=\int dx^3 J_A^t$ is not conserved and that its mean square change (e.g.~on a thermal ensemble) is related to that of the Chern-Simons number as
\be
\langle (\Delta Q_A)^2\rangle = q^2 \langle (\Delta N_{CS})^2\rangle\,.
\label{axan}
\ee 
In turn, using the anomaly relation, the definition of the Chern-Simons diffusion rate \eqref{CSdef} could be rewritten as
\begin{equation}
\frac{\Gamma_{CS}}{2T} = -\frac{1}{q^2}\lim_{\om\to0}{\rm Im}\left[\frac1{\om}  G_R^{DD}(\om,\vec{k}=0)\right]\,,
\end{equation}
where $G_R^{DD}(k)$ is the retarded Green's function of the divergence of the non-conserved axial current. Fourier transforming the term in square bracket back to position space we obtain
\begin{equation}
\frac1{\om} \lim_{\vec{k}\to0} G_R^{DD}(k) =\rmi\int\limits_{-\infty}^{+\infty} \rmd t \ex{\rmi \om t} \del_t\langle Q_A(t)\,Q_A(0) \rangle_R\,.
\end{equation}
Taking now the $\om\to0$ limit we get
\begin{equation}
\lim_{\om\to0}\frac1{\om} \lim_{\vec{k}\to0} G_R^{DD}(k) = \rmi\langle Q_A(t\to+\infty)\,Q_A(0) \rangle_R\,.
\end{equation}
Since the charge $Q_A$ is not conserved the above retarded correlator vanishes as $t\to\infty$. Therefore, in presence of a chiral anomaly, \eqref{CSdef} does not provide a good definition of the Chern-Simons diffusion rate. 

In other words, the modes associated to the non-conserved charge $Q_A$ are gapped and do not survive the hydrodynamic limit. Indeed, the equilibrium, time-independent value of a non-conserved charge necessarily vanishes and the decay in time of the charge is obscured by the limit. 

In order to study the explicit decay of the total chiral charge one should look at the retarded correlator $G^{tt}_R(\om)=\langle J_A^t J_A^t\rangle_R$ at zero momentum $\vec{k}=0$ and finite frequency $\om$. Generically, the retarded correlator can be written as a sum of poles plus an analytic, scheme-dependent part. In particular, for a non-conserved charge one expects the singular part of the correlator to be of the form\footnote{There might be additional poles and branch cuts in the correlators. We focus on the pole closest to the origin in the complex $\om$ plane, which is the one that dictates the late time behavior.}
\be\label{poles}
G_R^{tt}(\om) \sim \frac{\rmi R}{\om+\frac{\rmi}{\tau} }~,
\ee
with $R$ and $\tau$ real and constant. The above correlator models the gapped decay mode of the axial charge and, indeed, taking the Fourier transform of both sides one obtains
\be
\langle Q_A(t) Q_A(0)\rangle_R \sim R\,\theta(t)\ex{-\frac{t}{\tau}}~,
\ee
where $\tau$ is immediately identified with the axial relaxation time. 

Notice that the imaginary part of the Green's function (\ref{poles}) reads
\be
{\rm Im}G_R^{tt}(\om) \sim \om \frac{R}{\om^2+ \tau^{-2}}\,,
\label{imaginarypole}
\ee
 so that, if $q\neq0$, 
 \be
 \frac{1}{\omega}{\rm Im}G_R^{Q Q} = \frac{\omega}{q^2}{\rm Im}G_R^{tt}\sim \frac{1}{q^2}\om^2 \frac{R}{\om^2+ \tau^{-2}}\,.
 \ee
Hence, the $\omega\rightarrow0$ limit of the left hand side, which one would take to get the Chern-Simons diffusion rate in absence of anomalies, would thus give zero.

The above observations suggest that a correct way to define the Chern-Simons diffusion rate in presence of anomalies would be by means of a cut-off $t_*\ll \tau$ in the time integration entering the related correlator in (\ref{def}) \cite{Moore:2010jd,Iatrakis:2015fma}, whenever the microscopic time scales involved in the CS number fluctuation are much smaller than $\tau$. 

A nice hydrodynamical model (realized holographically in the Witten-Sakai-Sugimoto theory), for axial charge diffusion and relaxation can be found in \cite{Iatrakis:2015fma}. Taking into account the thermal fluctuations for the average squared axial charge, which at equilibrium are related to the axial susceptibility $\chi_A$, one can write, for $t\ll\tau$
\be
\langle (\Delta Q_A)^2\rangle \sim \chi_A T \left[ 1 - e^{-\frac{2t}{\tau}}\right] V \approx \frac{2\chi_A T}{\tau} V t \equiv q^2 \Gamma_{CS} V t\,,
\label{realt}
\ee
where $V$ is the spatial volume and in the last step we have used (\ref{axan}). 
Crucially, this expression can make sense if the Chern-Simons diffusion rate is that of the $q=0$ theory
\be
\Gamma_{CS}\equiv\Gamma_{CS}(q=0)\,.
\label{CSq0}
\ee
The above formulae imply, in turn, that the frequency gap, i.e.~the inverse relaxation time, is given by (see also \cite{Moore:2010jd} and appendix \ref{relax} for an alternative derivation) 
\be
\frac{1}{\tau} = \frac{q^2 \Gamma_{CS}}{2\chi_A T}\,.
\label{fenof}
\ee
Consistently, when $q\rightarrow0$, $\tau\rightarrow\infty$, hence the definition (\ref{CSq0}) can be read as the one which arises in the limit in which the cut-off in the time integration is sent to infinity.

This framework is akin to
the Witten-Veneziano formula 
\be
m_{WV}^2 = \frac{2N_f \chi_g}{f_{\pi}^2}\,,
\ee
which, in large $N$ QCD with $N_f$ massless flavors at $T=0$, gives the squared $\eta'$ mass (driven by the axial anomaly) in terms of the topological susceptibility $\chi_g$ (the Euclidean counterpart of $\Gamma_{CS}$) of the unflavored ($N_f=0$) theory (the topological susceptibility being zero for $q\neq0$ with massless flavors). 

As we are going to show in the following subsections, holography consistently implements the above observations. Moreover, it also allows, in principle, to work beyond the hydrodynamical limit and to extract the complete quasi-normal mode spectrum of the correlators. 
\subsection{The holographic approach}
The holographic dual of a 4d QFT with a $U(1)$ anomalous (axial) current includes a universal sector
described  by a Stueckelberg action \cite{Klebanov:2002gr}
\begin{equation}
S = -\frac1{4\ka^2_5} \int \rmd^5 x\sqrt{-g} \left[ \frac{a}2 |F_A|^2  + H\(\rmd C + qA\)^2 \right]\,,\quad F_A=\rmd A\,.
\label{stueck}
\end{equation}
We omit possible Chern-Simons terms since these are more than quadratic in the vector field $A$ and cannot contribute to the two-point function of the dual operator. 
The parameter $q$ is constant while the kinetic coefficient $a$ and the mass $H$ are in general functions of background fields. 
The action (\ref{stueck}) is found in all the dimensional reduction to five dimensions of top-down holographic models with axial anomaly, $q$ being precisely the anomaly coefficient (see e.g.~\cite{Klebanov:2002gr,Cotrone:2012um,Iatrakis:2015fma} and appendix \ref{appendixmn}).

Here we will treat $A$ and $C$ as fluctuating fields over a fixed background. 
In the simplest case where the metric is the only background field, the coefficients will all be constant and can be set to any desired value rescaling the fields $A$ and $C$. 
However, we will keep them unfixed to find a more general expression for the equations of motion.

The action above is invariant under the transformation
\begin{equation}
\d A = \rmd\l\,,\qquad \d C = -q\,\l~,
\end{equation}
which, enforced at the boundary on the couplings $\int A_\mu J_A^{\mu} + C\, Q$, precisely implies the operator relation 
\begin{equation}
\del_\m J_A^\m + q\,Q\simeq0
\end{equation}
in the dual field theory, where $J_A$ is the current dual to the vector $A$ and $Q$ the operator dual to the scalar $C$.

The combination $B=\rmd C +q A$ is invariant under the above local transformation. In addition we have $\rmd B = q\,\rmd A$, so when $q\neq0$ the action can be rewritten in terms of $B$ only
\begin{equation}
S = -\frac1{4\ka^2_5\,q^2} \int \rmd^5 x \sqrt{-g} \left[ \frac{a}2 |F_B|^2  + q^2H\,B^2 \right]\,.
\label{actionB}
\end{equation}
In this formulation the holographic map reads
\begin{equation}
B_\m \leftrightarrow J_B^\m = \frac1{q}J_A^\m\,,
\end{equation}
so that we have
\begin{equation}\label{anomalyrel}
\del_\m J_B^\m + Q\simeq0
\end{equation}
inside correlation functions. In particular, the retarded correlator of the operator $Q$ satisfies
\be\label{anomalyrel}
\langle Q(x) \, Q(0) \rangle_R = \langle \del_\m J_B^\m(x) \del_\n J_B^\n(0) \rangle_R~.
\ee
\subsection{Equations of motion}
Let us consider a 5d background metric of the form
\be
ds^2 = g_{\mu\nu}dx^{\mu}dx^{\nu} + g_{uu}du^2\,,
\label{genb}
\ee
where the metric components are taken to be functions of the radial variable $u\in[u_h, u_0]$. The metric is assumed to have an horizon at $u=u_h$ and to be asymptotically (i.e.~at $u\rightarrow u_0$) described by radial slices of Minkowski 4d spacetime.  Notice the crucial assumption $g_{u\m}=0$.
We denote as $M$ the five-dimensional index, $\mu,\nu$ the four dimensional ones, $i,j$ the three spatial directions and $k=(\omega, \vec k)$.  

Since eventually we will compute Green's function in momentum space, it is convenient to use the Fourier transformed fields from the beginning. Defining
\begin{equation}
B_M(x,u) \equiv \int \frac{\rmd^4 k }{(2\pi)^4} \tilde B^k_M(u) \ex{-\rmi k\cdot x}\,,
\end{equation}
and substituting inside the action (\ref{actionB}) we find (dropping tildes)
\begin{align}
S &= -\frac1{4\ka^2_5\,q^2} \int \frac{\rmd^4 k }{(2\pi)^4}\int \rmd u \sqrt{-g} \left\{ag^{uu}g^{\m\n}B_\m^{\prime-k}B_\n^{\prime k} + \rmi ag^{uu}\hat k^\m B_\m^{\prime -k}B_u^{k} + \right.  \nn
&\left. - \rmi ag^{uu}\hat k^\m B_u^{-k}B_\m^{\prime k}+B_\m^{-k}\left[\(a\hat k^2+q^2H\) g^{\m\n}- a\hat k^\m\hat k^\n\right]B_\n^k +\right.  \nn 
&\left.+ \(a\hat k^2+q^2H\)g^{uu}\,B_u^{-k}B_u^k \right\}\,,
\end{align}
where $'$ denotes derivative along the radial direction and hatted quantities are contracted using the metric $g_{\m\n}$.
The equations of motion resulting from varying with respect to $B_\m^{\prime-k}$ are
\begin{equation}
\(a\,\sqrt{-g}g^{uu} g^{\m\n} B^{\prime k}_\n\)' + \rmi \(a\,\sqrt{-g} g^{uu} \hat k^\m B_u^k\)' -\sqrt{-g} \left[\(a\hat k^2+q^2H\) g^{\m\n}- a\hat k^\m\hat k^\n\right]B_\n^k=0\,,
\label{prev}
\end{equation}
whereas varying with respect to $B_u^{-k}$ yields
\begin{equation}
  \(a\hat k^2+q^2H\)B_u^{k}=\rmi a \hat k^\m B_\m^{\prime k}\,.
\end{equation}
The above equation can be used to remove $B_u^k$ from (\ref{prev}), thus obtaining
\begin{equation}\label{eomI}
\(a\,\sqrt{-g}g^{uu} K^{\m\n} B^{\prime k}_\n\)'  -\sqrt{-g} \(a\hat k^2+q^2H\) K^{\m\n} B_\n^k=0\,,
\end{equation}
with
\begin{equation}
K^{\m\n} \equiv  g^{\m\n} - a\frac{\hat k^\m\hat k^\n}{\(a\hat k^2+q^2H\)}\,,
\end{equation}
which for $q\to0$ becomes the projector onto the subspace transverse to $k^\m$. From (\ref{eomI}) and its counterpart for $B_\m^{-k}$, we find that 
\begin{equation}\label{conserved}
\frac{\rmd}{\rmd u} \left[ a\sqrt{-g}g^{uu}K^{\m\n}\(B^{-k}_\m B_\n^{\prime k} - B^{k}_\m B_\n^{\prime -k}\) \right] = 0\,,
\end{equation}
so the quantity in parenthesis is {\em conserved along radial motion}.

The on-shell action reads
\begin{equation}\label{on-shell}
S_\text{on-shell} = -\frac1{4\ka_5^2\,q^2} \left. \int \frac{\rmd^4 k }{(2\pi)^4} B_\m^{-k} \(a\,\sqrt{-g}\,g^{uu} K^{\m\n}\)  B_\n^{\prime k}\right|_{u_h}^{u_0}~.
\end{equation}

\subsection{Retarded Green's functions}
Thanks to the anomaly relation \eqref{anomalyrel}, the longitudinal Green's function of the current operator $J_B$ is equal to the retarded Green's function of the topological operator $Q$
\begin{equation}
G_R^{LL}(k)\equiv -k_\m k_\n G_R^{\m\n}(k)=-k_\m k_\n\langle J_B^\m(-k) J_B^\n (k)\rangle_R = \langle Q(-k) Q (k)\rangle_R~.
\label{GLL}
\end{equation}
The Chern-Simons diffusion rate is defined in terms of the retarded Green's function of the topological operator $Q$ as follows
\begin{equation}\label{CSdefinition}
\Gamma_{CS} = - 2\,T\lim_{\om\to0}\lim_{{\vec k}\to0} \frac{{\rm Im}[\langle Q(-k) Q (k)\rangle_R]}\om
			= - 2\,T\lim_{\om\to0}\lim_{{\vec k}\to0} \frac{{\rm Im}[G_R^{LL}(k)]}\om ~.
\end{equation}
The holographic computation of the Green's functions is formally done through the following steps: a) setting $B^k_{\mu}(u)= b_{\mu}^{\nu\, k}(u) b^k_{\nu}$, where $b^k_{\nu}= k_{\nu}$ can be chosen to implement the boundary condition $B_{\mu}\rightarrow\partial_{\mu} C$ in momentum space (both $b_{\mu}^{\nu\, k}(u)$ and $C$ are taken to be normalized to one at the boundary); b) taking the functional (second) derivative of the on-shell action (\ref{on-shell}) with respect to $b^{-k}_{\mu}$ (and $b^{k}_{\nu}$).  Setting $\vec k=0$, with $b_t^{t\, \omega}\equiv b_t^{\omega}$ (and zero for the other components) we get
\be
{\rm Im}[G_R^{LL}(\omega,\vec 0)] =-\frac{\rmi \omega^2}{4\ka_5^2\,q^2}\lim_{u\to u_h}\left[ a\sqrt{-g}g^{uu}K^{tt}\(b^{-\omega}_t b_t^{\prime\,\omega} - b^{\omega}_t b_t^{\prime\,-\omega}\) \right]\,,
\label{corre}
\ee
where the right hand side, being independent on the radial variable (see (\ref{conserved})), can be equally computed at the horizon (as we do above) or at the boundary (as according to the standard holographic prescription \cite{Son:2002sd}). This allows to deduce the generic behavior of the Green's function just by considering universal features of the near-horizon solution. 

In order to compute (\ref{corre}), we need to extract the generic near-horizon 
behavior of a massive vector field in the background (\ref{genb}). We will just make use of the following assumptions:
a) the metric is diagonal and depends on the radial coordinate $u$ only; b) the $uu$ component has a simple pole at $u=u_h\equiv1$ (we suitably rescale our coordinates so that the latter equivalence holds); c) 
the $tt$ component has a simple zero at $u=1$; d) the $ii$ components are proportional to the identity and finite at $u=1$; e) the mass of the vector is finite at $u=1$; f) the kinetic term of the vector is finite at $u=1$.

Let us consider the equation of motion for the $t$-component of the vector field in momentum space setting $k_i=0$
\be\label{eomgeneric}
\partial_u \left(\frac{\sqrt{-g}\, a\, H g^{tt} g^{uu}}{Hq^2+ a\, \omega ^2 g^{tt}}\partial_u b^{\omega}_t\right)=\sqrt{-g} H g^{tt} b_t^{\omega}\,,
\ee
and consider the following expansions
\bea
&& a(u) = \sum_{n=0}^{\infty} a_h^{(n)}(1-u)^n\,,\quad H(u) = \sum_{n=0}^{\infty} H_{h}^{(n)}(1-u)^n\,,\quad g^{xx} = \sum_{n=0}^{\infty} c_{x}^{(n)}(1-u)^n\,,\nonumber \\
&& g^{tt} = - (1-u)^{-1}\sum_{n=0}^{\infty} c_{t}^{(n)}(1-u)^n\,,\quad g^{uu} = (1-u) \sum_{n=0}^{\infty} c_{u}^{(n)}(1-u)^n\,, \nonumber \\
&& b_t^{\omega}(u) = (1-u)^{\alpha}\sum_{n=0}^{\infty}b_{h}^{(n)} (1-u)^n\,, 
\eea
with $c_{t}^{(0)}>0, c_{x}^{(0)}>0, c_{u}^{(0)}>0$. Solving the equation of motion (\ref{eomgeneric}) and choosing the incoming wave solution at the horizon sets
\be
\alpha = -\rmi \omega \sqrt{\frac{c_{t}^{(0)}}{c_u^{(0)}}}\,.
\ee
Moreover one finds
\be
\frac{b_{h}^{(1)}}{ b_{h}^{(0)}}=-\rmi \frac{q^2}{\om}\frac{  H_{h}^{(0)}}{a_{h}^{(0)}\sqrt{c_{t}^{(0)}c_{u}^{(0)}}}+\frac{q^2  H_{h}^{(0)}}{a_{h}^{(0)}c_{u}^{(0)}}+ {\cal O}(\omega)\,.
\label{bh1}
\ee
Now the point is that the Chern-Simons diffusion rate as defined in (\ref{CSdefinition}) turns out to be given by \begin{equation}
\Gamma_{CS} = \frac{T}{\ka_5^2} \frac{H_h^{(0)}}{(c_x^{(0)})^{3/2}}| b_h^{(0)}|^2\,.
\end{equation}
In the $q=0$ case, $|b_h^{(0)}|=1$ reproduces the relation (\ref{general}). Crucially, however, in the generic case equation (\ref{bh1}) holds. Since we are interested in the $\omega\rightarrow0$ limit, we can just look at the first term in that equation. From this we immediately see that if we want $b_{h}^{(0)}$ to be finite, then $b_{h}^{(1)}\rightarrow\infty$ in the limit. Otherwise, if we want to avoid this singular behavior keeping $b_{h}^{(1)}$ finite, we need to assume that
\be
b_{h}^{(0)}\sim\omega\rightarrow0\,,
\ee
so that, consistently with the non-conservation of the axial charge, the Chern-Simons diffusion rate as defined in (\ref{CSdefinition}) {\it vanishes.}

We have checked this general result in the simplest setup where  $A$ and $C$ are treated as fluctuations over a fixed Schwarzschild-AdS background (see appendix \ref{adsb} for details). 
Solving the related equations of motion numerically, the analysis confirms that $b_h^{(0)}=0$ whenever $q\neq0$ as shown in figure 1.
Note that for $q \to 0$ but non vanishing, $|b_h^{(0)}(\omega)|^2$ approaches the behavior of the $q=0$ case for $\omega >0$, but eventually it always drops to zero at $\omega=0$.
\begin{figure}[ht!]
\centering
\includegraphics[scale=.7]{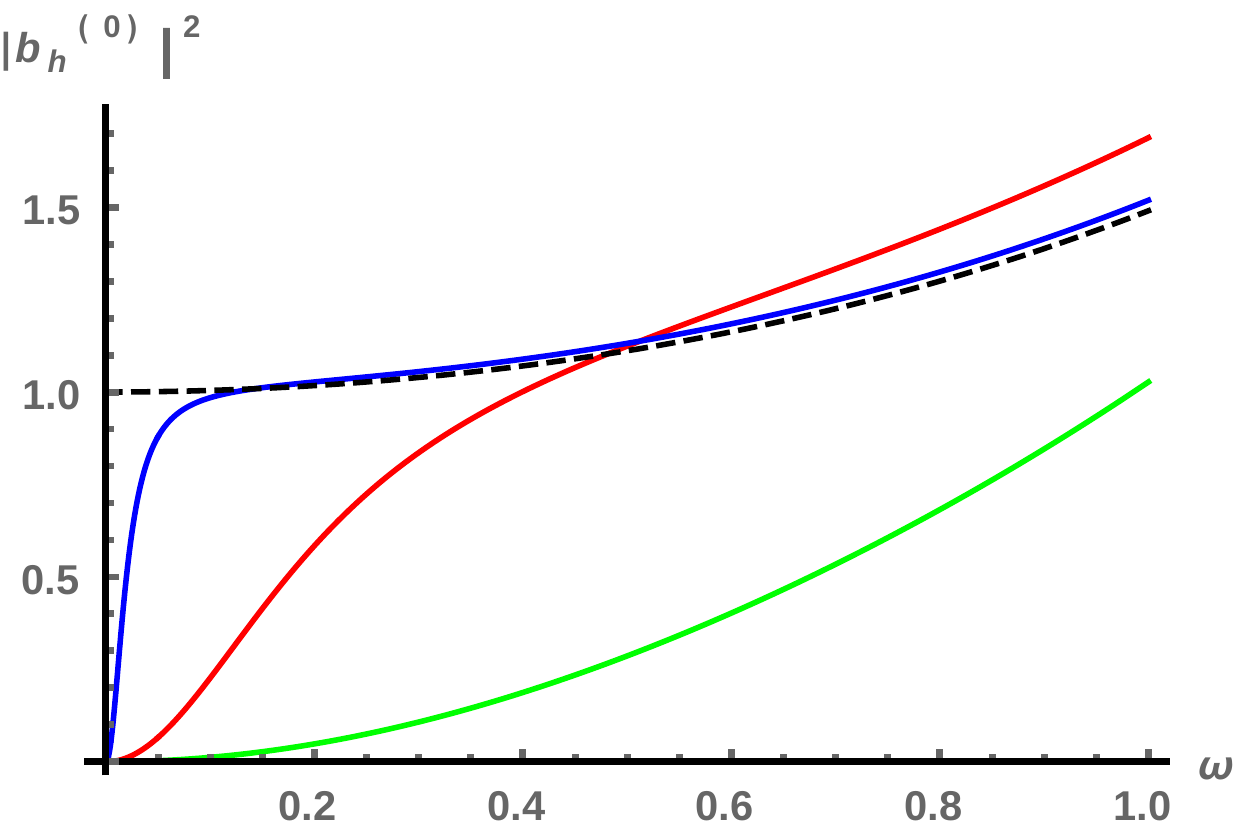}
\caption{The coefficient $|b_h^{(0)}|^2$ as a function of $\omega$ for an AdS$_5$ black hole background with unit radius and $a=H=1$ for different values of $q=0.04\,, 0.44\,,3$ (top to bottom at small $\omega$) and for $q=0$ (dashed line).
}
\label{logDipolezmfig}
\end{figure}

\subsection{Quasinormal modes}

Holography implies that the location of the poles of a two-point retarded correlator are given by the frequencies of the quasinormal modes of the corresponding bulk field \cite{Danielsson:1999zt,Son:2002sd}, so their study allows to extract the relaxation time $\tau$.
The quasinormal mode spectrum can be computed in the following way.

Let $\Phi$ be the bulk field dual to an operator $\mathcal O$ in the field theory, and let $\Phi$ fluctuate over a background with an event horizon. 
The fluctuations satisfy second order linear equations of motion whose solution near the horizon will be the sum of incoming and outgoing waves. 
As above, since we are after the retarded correlator, we pick the ingoing solution. 
The solution close to the boundary, placed at $u=0$,  will have the form
\be
\Phi(u,\om,\vec{k}) = A(\om,\vec{k})\,u^{\Delta_-} + B(\om,\vec{k})\,u^{\Delta_+}\,,
\ee
where $A$ and $B$ are completely determined up to a common non vanishing multiplicative factor $f(\om,\vec{k})\neq0$. 
Quasinormal modes are solutions satisfying ingoing boundary conditions at the horizon that vanish at infinity. 
Their frequencies, $\om_n(\vec k)$, are thus defined implicitly by the equation
\be\label{qnmdef}
A(\om_n(\vec k), \vec{k}) = 0~.
\ee
In order to see the connection with the poles of the retarded correlator, recall that the correlator is \cite{Son:2002sd} 
\be
\langle\mathcal O\,\mathcal O\rangle_R \sim \frac{B}{A} + \dots
\ee
where $\dots$ stand for scheme dependent contact terms. The zeros of $A$ are exactly the poles of the correlator. According to \eqref{imaginarypole} we need to find the first zero of
\be
\left( {\rm Im}\frac{B(\om)}{A(\om)}\right)\inv
\ee
along the imaginary $\om$ axis. Applying this logic to the AdS black hole background, one finds \cite{Jimenez-Alba:2014iia} that the axial relaxation time $\tau$ scales like $q^{-2}$ for small $q$. Since $q$ enters the equations of motion only as $q^2$ even in the most general case \eqref{eomgeneric}, we expect the relation $\tau\sim q^{-2}$ to be universal for small non-zero $q$. This relation would thus reproduce the phenomenological formula (\ref{fenof}).

\vskip 15pt \centerline{\bf Acknowledgments} \vskip 10pt \noindent We are grateful to Riccardo Argurio, Matthias Kaminski, Dmitri Kharzeev, Karl Landsteiner, Javier Mas, Daniele Musso and Javier Tarrio for relevant comments and discussions. ALC is partly supported by the Florence University grant ``Fisica dei plasmi relativistici: teoria e applicazioni moderne".



\appendix

\section{Chern-Simons diffusion rate in the ${\cal N}=4$ plasma with external magnetic field}\label{appenkhar}
\setcounter{equation}{0}

The effective 5d holographic description of the ${\cal N}=4$ SYM plasma in presence of a constant magnetic field is provided by a magnetically charged black hole solution arising from the Einstein-Maxwell-Chern-Simons action
\be
S= -\frac{1}{16\pi G_5}\int d^5x\sqrt{-g}\left[ R + F^{MN}F_{MN} -\frac{12}{L^2}\right] + \frac{1}{6\sqrt{3}\pi G_5}\int A\wedge F \wedge F\,,
\ee
supplemented by standard boundary terms. In units $L=1$, the solution follows from the ansatz (see e.g. \cite{Basar:2012gh} and references therein)
\bea
ds^2 &=&  -U(r) dt^2 + \frac{dr^2}{U(r)} +\frac{e^{2 V(r)}}{v}(dx_1^2 + dx_2^2)+\frac{e^{2W(r)}}{w} dx_3^2\,,\nonumber \\
F &=& \frac{B}{v} dx_1\wedge dx_2\,,
\eea
where $U,V,W$ are functions of the radial variable $r$, the coordinates are rescaled in such a way that the black hole horizon is at $r=1$ where $U(1)=0, U'(1)=1, V(1)=W(1)=0$ and $v,w$ are functions of the magnetic field $B$.
The Chern-Simons diffusion rate for the dual magnetized plasma has been computed in \cite{Basar:2012gh} finding
\be
\Gamma_{CS} = \frac{g_{YM}^4}{2^7 \pi^5} \frac{N^2}{2\pi} \frac{T}{v\sqrt{w}}\,,
\ee
where $T$ is the temperature ($T=1/4\pi$ in the rescaled coordinates defined above). It is easy to realize that this formula, which holds for any value of the magnetic field, precisely matches our eq.~(\ref{result}), since the black hole entropy density is given by
\be
s = \frac{A_h}{4 V_3 G_5} = \frac{N^2}{2\pi}\frac{1}{v\sqrt{w}}\,,
\ee
where $A_h$ is the area of the horizon, $V_3$ is the (infinite) 3d-space volume and the holographic relation $G_5 = \pi/2 N^2$ has been used. 

\section{Chern-Simons diffusion rate in the anisotropic ${\cal N}=4$ plasma}\label{appenmat}
\setcounter{equation}{0}
The anisotropic ${\cal N}=4$ SYM plasma considered in \cite{matranca} corresponds to thermal ${\cal N}=4$ SYM deformed by a linear space-dependent topological theta angle
\be
\theta(x_3) = 2\pi a x_3\,,
\ee
where $x_3$ is one of the space directions and $a$ is a dimensionful anisotropic parameter which can be read as the density of homogeneously smeared D7-branes along the $x_3$ direction. The dual black hole solution arises from the effective 5d action (Einstein frame)
\be
S = \frac{1}{16\pi G_5}\int d^5 x\sqrt{-g}\left[R + 12 - \frac{(\partial\phi)^2}{2}-e^{2\phi}\frac{(\partial C_0)^2}{2}\right]\,,
\ee
where $\phi$ is the dilaton and $C_0$ is the axion dual to the topological charge density operator of the QFT. The background has a metric of the form
\be
ds^2 = \frac{e^{-\phi(u)/2}}{u^2}\left(- {\cal F}(u){\cal B}(u) dt^2 + \frac{du^2}{{\cal F}(u)} + dx_1^2 + dx_2^2 + {\cal H}(u) dx_3^2\right)\,,
\ee
supported by a running dilaton $\phi(u)$ and the axion $C_0=a x_3$. The horizon is at $r=r_h$. The Chern-Simons diffusion rate, as computed in \cite{bu}, has, for any value of $a/T$, precisely the same form given in (\ref{result}), provided we recall that in this case
\be
g_{YM}^2(T) = 4\pi e^{\phi(r_h)}\,.
\ee

\section{Chern-Simons diffusion rate in WYM}\label{appendixwym}
\setcounter{equation}{0}

In this appendix we review the calculation of the Chern-Simons diffusion rate in Witten's Yang-Mills model \cite{Witten:1998zw} and show that it obeys the universal formula (\ref{result}).
The diffusion rate has been calculated in \cite{ben} (in string frame) and reads
\be
\Gamma_{CS} = 4\frac{1}{2\pi}\frac{\lambda_4^3}{3^6 \pi^2}(2 \pi R_4)^2 T^6\,,
\ee
where the overall factor $4$ (instead of the $1$ in \cite{ben}) originates from the difference between our conventions (\ref{identifications}) and (\ref{realC}) and the corresponding ones adopted in \cite{ben}.  In this formula $T$ is the temperature, 
\be
\lambda_4= 2\pi g_s N l_s M_{KK}=\frac12 g_{YM}^2 N \,,
\ee
is proportional to the ``UV 't Hooft coupling'' and $M_{KK}=1/R_4$ is the mass scale of the theory.
Since the entropy density is
\be
s= \frac{2^8 \pi^4 \lambda_4 N^2 T^5}{3^6 M_{KK}^2}\,,
\ee
one can write
\be\label{Gwym}
\Gamma_{CS} = \frac{g_{YM}^4}{2^7\pi^5}\, s\, T\,,
\ee
which is exactly in the universal form (\ref{result}).

It is instructive to see how the result above is generated in Einstein frame.
The five dimensional reduction of the WYM background has been performed in \cite{benincasa}.
The gravitational constant reads
\be
\frac{1}{2\kappa_5^2} = \frac{1}{2\kappa_{10}^2} 2\pi R_4 {\rm Vol}(S_4) = \frac{1}{2\kappa_{10}^2} \frac{16\pi^3}{3 M_{KK}}\,,
\ee
where $R_4$ is the radius of the circle of the cigar and ${\rm Vol}(S_4)$ is the volume of the unit radius four sphere.

The reduction of the term of the ten dimensional action containing $C_1$, whose integral on the circle (with coordinate $x_4$) is dual to the operator $Q$ (up to a constant),
\be
\frac{1}{2\kappa_{10}^2} \int d^{10}x \sqrt{-g_{10}}\, e^{\frac32 \phi}\, F_2^2\,,
\ee
gives
\be\label{redactionwym}
\frac{1}{2\kappa_{5}^2} \int d^{5}x \sqrt{-g_{5}}\, e^{\frac32 \phi -2f -8 w}\, F_1^2\,,
\ee
where $F_1 = dC_{x_4}$.
The functions $f, w$ are the metric functions in the reduction ansatz
\be
ds^2_{10} = e^{-\frac{10}{3}f} ds^2_5 + e^{2f}\left[e^{8w}dx_4^2 + e^{-2w}dS_4^2\right]\,.
\ee
Amazingly enough, on the background\footnote{We use the convention in which $g_s$ is in the brane tension and in $\kappa_{10}$, so there are no factors of $g_s$ in the dilaton solution.} 
\be\label{combination}
e^{\frac32 \phi -2f -8 w} = 1\,.
\ee
Thus, the five dimensional action is the one of a minimally coupled scalar.\footnote{Referring to the general action (\ref{actilde}), we have $e^{2\phi}=g_{44}$.} 
Considering that the field dual to $Q$ is not $C_1$ but $\frac{2 \pi R_4}{g_s l_s} C_1$ after reduction on the circle of the D4-brane action, as shown in \cite{ben}, 
one obtains that (\ref{redactionwym}), (\ref{combination}) give the result (\ref{Gwym}) for the CS diffusion rate in the WYM theory.

A remark is in order.
While the WYM has a running coupling in the confined phase, the gravity combination of fields above dual to the coupling in the deconfined case is again a constant.
This is due to the fact that the theory in the deconfined case is basically a six dimensional CFT.
In fact, one can calculate the four dimensional coupling from the action of a probe D4-brane wrapped around the circle of the cigar, as in \cite{noiwym}, obtaining in the deconfined phase
\be
\frac{1}{g^2_{YM,dec}} = \frac{1}{8 \pi^2 l_s} \int dx_4 e^{-\phi} \sqrt{g_{44}} = \frac{\beta_4}{8 \pi^2 l_s g_s}\,,
\ee
where $\beta_4$ is the length of the circle.
Thus, the 't Hooft coupling is constant and equals precisely $2\lambda_4$,
\be
g^2_{YM,dec} N = 2\lambda_4.
\ee

\section{Chern-Simons diffusion rate in MN}\label{appendixmn}
\setcounter{equation}{0}

In this appendix we calculate the Chern-Simons diffusion rate in the Maldacena-Nu\~nez (MN) model of ${\cal N}=1$ Super Yang-Mills \cite{Chamseddine:1997nm,mn} and show that it obeys the universal formula (\ref{result}).
The non-extremal solution corresponding to the deconfined, chirally symmetric phase of the dual ${\cal N}=1$ SYM theory is a linear dilaton background \cite{gtv}, i.e.~the dual of the LST representing the UV completion of the model.
The internal metric is the product of a two sphere and a three sphere.
In Einstein frame the solution reads
\bea \label{gtvsol}
&& ds^2_E = e^{\phi/2} \left[ -\nu(r) dt^2 + dx_i dx^i + \nu(r)^{-1} dr^2 + dS_2^2 + dS_3^2 \right]\,, \\
&& F_3 = P \left[ \Omega_3 + w_3 \wedge \Omega_2 \right]\,, \\
&& \phi \sim r\,.
\eea
In this formulae $P$ is proportional to the number of colors, $\Omega_{2,3}$ are the two and three-sphere volume forms and $w_3 = d\psi + \cos{\theta_2} d\phi_2$ is the third left-invariant one-form of $SU(2)$ (the other two will be denoted as $w_{1,2}$).
The two and three spheres have radii equal to 2.

The identification of the gravity field dual to $Q$ has been done directly in \cite{mn}, see also \cite{divecchia,bertolini,noi}.
It is a component of the RR two-form potential $C_2$, whose UV asymptotics give, upon integration on a two-sphere, the $\theta_{YM}-$term,
\be
Q \leftrightarrow q(r) \left(w_1 \wedge w_2 - \Omega_2\right) \equiv q(r) \Phi \,,  \qquad {\rm with}\quad q(r) \rightarrow \psi - \psi_0 \quad {\rm for} \quad r \rightarrow \infty\,.
\ee
Note that $\psi$ is a coordinate of the three-sphere, while $\psi_0$ is a constant proportional to $\theta_{YM}$.

The five dimensional consistent reduction of type IIB containing the relevant fields for our purposes can be found in \cite{cassani}. 
The relevant component of $C_2$ is termed $c^{\Phi}$.
The full action is quite involved.
But we are interested only in the quadratic fluctuations around the solution (\ref{gtvsol}).
It can be checked that the only terms contributing are in the form of the usual Stueckelberg coupling of $c^{\Phi}$ with the Reeb vector $A$.
Let us sketch the derivation.

Comparing the metric (3.3) in \cite{cassani} with (\ref{gtvsol}) we get that the only metric fields which are turned on are
\be
v=\frac{\phi}{4} + \log{3}\,, \qquad u=\frac{\phi}{4} + \frac12 \log{6}\,, \qquad g_{\mu\nu} = \frac{1}{9 \cdot 2^{4/3}} e^{\frac43 \phi} {\rm diag}(-\nu, 1,1,1,\nu^{-1}) \,. 
\ee
Apart from the dilaton, the other quantity in \cite{cassani} which is different from zero is the flux $q=18 P$.

The only five dimensional field where $c^{\Phi}$ enters, which is linear in the fluctuating fields, is 
\be
g_1^{\Phi} = d c^{\Phi} - q A \,.
\ee
In the action it enters in a quadratic term and some mixing terms with other fields, with specific coefficients depending on the metric. 
On the background the only non-zero coefficient is the one of the quadratic term (which is 1), so the action reduces to\footnote{We correct a factor of $\frac12$ in \cite{cassani}.}
\be\label{actiong}
S_c = - \frac{1}{2 \kappa_5^2} \int \frac{1}{6^2} \frac12 (g_1^{\Phi})^2 \star 1 \,.
\ee

One can also check that $A$ appears always in terms which are at least of third order in fluctuating fields, apart from its kinetic term
\be
S_A = - \frac{1}{2 \kappa_5^2} \int \frac{1}{2} (54)^{4/3} e^{\frac43 \phi} (d A)^2 \star 1 \,.
\ee

Now, by comparing formula (2.49) in \cite{divecchia} with the notation in \cite{cassani} one gets that the field dual to $Q$ is\footnote{We work in units $\alpha'=1$ here.}
\be
Q \leftrightarrow C(r) \equiv \frac{8 \pi^2}{g_{YM}^2} \left( \frac{c^{\Phi}}{6}\right)\,. 
\ee
Thus, if the effects of anomaly are ignored, i.e.~we drop the terms in $A$, the action (\ref{actiong}) reads
\be\label{ac5mn}
S_c = - \frac{1}{2 \kappa_5^2} \int \sqrt{g_5} \left(\frac{g_{YM}^2}{8\pi^2} \right)^2 \frac{1}{2} (dC)^2\,,
\ee
which is in the form (\ref{acC}), implying that the CS diffusion rate is in the universal form (\ref{result}).

As an aside, note that the function $H$ of formula (\ref{generalac}) is a constant in the MN model.\footnote{Referring to the notation in (\ref{actilde}), $e^{2\phi}=\det(g_{\Omega_{2}'})$ and we get (\ref{actiong}) once we take into account the $\frac16$ normalization in \cite{cassani}.}
The latter has no $AdS$ UV asymptotics but in the deconfinement phase the coupling is anyway constant.
In fact, if we calculate it as in formula (4.2) of \cite{divecchia}, but on the non-extremal solution, we get 
\be
\frac{1}{g^2_{YM,dec}} = \frac{1}{4 \pi^2 g_s}\,.
\ee
\section{Notes on the axial relaxation time}\label{relax}
\setcounter{equation}{0}
In this appendix a quick derivation of formula (\ref{fenof}) is presented.
Let us turn off the anomaly for a moment. Diffusion of the axial current implies Fick's law
\be
\vec J_A = - D\vec\nabla J^t_A\,,
\ee
where $D$ is the diffusion constant. In momentum space
\be
J_A^x = - \rmi k D J_A^t\,,
\label{rel1}
\ee
where $k=k_x$. Now, turning on a source $A_x$ for $J_A^x$ we have, in linear response,
\be
\langle J_A^x \rangle = - G_R^{xx} A_x\,,
\ee
where $G_R^{xx}$ is the retarded correlator of $J_A^x$. At zero spatial momentum, considering the electric field
\be
E_x = -\rmi \omega A_x\,,
\ee
the previous relation reproduces Ohm's law with
\be
G_R^{xx}(\omega, \vec 0) = \rmi\omega \sigma\,,
\ee
where $\sigma$ is the conductivity.

Using (\ref{rel1}) we thus get
\be
\langle J_A^t\rangle = \frac{\langle J_A^x\rangle}{-\rmi k D} = \frac{G_R^{xx}}{\rmi k D} A_x\,.
\ee
Now, the anomaly, through the gauge invariant combination $\partial_{\mu} C + q A_{\mu}$, implies that a source $C$ for the topological charge density $Q(x)$, is induced by the electric field as
\be
\rmi k C = q A_x\,,
\ee
so that
\be
\langle Q\rangle = - G_R^{QQ} C = -\frac{G_R^{QQ}}{\rmi k} q A_x = \frac{\Gamma_{CS}\, \omega}{2k T} q A_x\,,
\ee
where we have used
\be
G_R^{QQ} = -\rmi\frac{\Gamma_{CS}}{2T}\omega + {\cal O}(k^2)\,.
\ee
All in all, from
\be
\frac{\langle Q\rangle}{\langle J_A^t\rangle}=\frac{\rmi\,\omega\, q\,\Gamma_{CS} D}{2T G_R^{xx}}=\frac{q\,\Gamma_{CS}\,D}{2T\sigma}+{\cal O}(k) \equiv \frac{1}{q\tau}+{\cal O}(k)\,,
\ee
and using $D/\sigma= \chi^{-1}_A$ we get the desired relation
\be
\frac{1}{\tau} = \frac{q^2 \Gamma_{CS}}{2T\chi_A}\,.
\ee
\section{Stueckelberg action on the AdS-BH background}\label{adsb}
\setcounter{equation}{0}
In the simplest possible setup the fields $A$ and $C$ defined in (\ref{stueck}) are treated as fluctuations over a fixed Schwarzschild-AdS background
\begin{equation}
\dsq = \frac{\ell^2}{z^2}\left[ -b(z)\,\rmd t^2 + |\rmd x|^2 + \frac{\rmd z^2}{b(z)}\right]\,,\qquad b(z) = 1- \frac{z^4}{z_h^4}\,,
\end{equation}
where $0 \leq z\leq z_h$, the horizon is at $z=z_h $ and the black hole temperature is given by $4\pi T =| b'(z_h)| = 4/z_h$.
Another useful description is obtained using the dimensionless radial coordinate $u\in[0,1]$
\begin{align}
&u = \frac{z^2}{z_h^2}\,,	\nn
& \dsq = \frac{\ell^2}{u}\left[ \frac{-b(u)\,\rmd t^2 + |\rmd x|^2}{z_h^2} + \frac{ \rmd u^2}{4u\,b(u)}\right]\,,\qquad b(u) = 1- u^2\,.
\end{align} 
Let us now solve the equation of motion (\ref{eomI}) on the above AdS-BH background.
\subsection{Near-boundary expansion}
Close to the boundary at $u=0$ the equation of motion \eqref{eomI} becomes (assuming $H$ and $a$ to go to 1 at the boundary)  
\begin{equation}
4u^2 B^{\prime\prime k}_\m - (q\ell)^2 B_\m^k \simeq 0\,,
\end{equation}
therefore we have the following expansion close to the boundary
\begin{align}\label{rescaled}
& B_\m^k(u) \equiv u^\Xi \,  f_\m^{k\,\n}(u)\, b_{\n}^k\,,		\nn
& f_\m^{k\,\n}(u) = \d_\m^\n + \cO(u) +u^{1-2\Xi} \( b_2^k \d_\m^\n + \cO(u)\)\,,
\end{align}
where the exponent $\Xi = \frac12\(1-\sqrt{1+(q\ell)^2}\)$ is strictly negative.
\subsection{Near-horizon expansion}
Close to the horizon at $u=1$ (again assuming $H$ and $a$ to go to constant values $H_h$, $a_h$) we have
\begin{equation}
K^{\m\n}\left[ 16(1-u)\((1-u)B^{\prime k}_\n\)' + w^2\, B^k_\n \right]=0\,,
\end{equation}
where $w=z_h\om$ and the matrix $K$ is evaluated at the horizon:
\begin{align}
& K^{ij} \to \frac{z_h^2}{\ell^2}\, \d^{ij}\,,			\nn
& K^{it} \to \frac{z_h^2}{\ell^2} \frac{k^i}\om	\,,	\nn
& K^{tt} \to \frac{(q\ell)^2H_h + a_h z_h^2 |\vec k|^2}{a_h\ell^2\om^2}~.
\end{align}
The matrix is non-singular at the horizon, and the near-horizon solution has the form
\begin{equation}
f_\m^{k\,\n}(u) = f_{h}^k{}_\m{}^{\n} \(1-u\)^{-\rmi \frac{w}4} \left[1+\cO(1-u)\right] + g_{h}^k{}_\m{}^{\n} \(1-u\)^{+\rmi \frac{w}4} \left[1+\cO(1-u)\right]\,.
\end{equation}
We pick the in-falling solution which corresponds to setting $g_{h}^k{}_\m{}^{\n}=0$.
\subsection{Retarded Green's functions and Chern-Simons diffusion rate}
In order to define the retarded Green's function for our current operator, $J_B$, we need to evaluate the action on a generic solution of the equations of motion. 
Since we are after the imaginary part of the correlator, we do not need to renormalize explicitly the action, the relevant part being finite. 
From \eqref{on-shell} and \eqref{rescaled} we have
\begin{equation}
S_\text{on-shell} = -\frac1{4\ka_5^2\,q^2} \left. \int \frac{\rmd^4 k }{(2\pi)^4}  b^{-k}_\m\left[ \sqrt{-g}\,a\,g^{uu}  u^{2\Xi-1}f_\r^{-k\,\m}K^{\r\s}\(\Xi f_\s^{k\,\n} + u f_\s^{\prime k\,\n}\)\right]\right|_0^1 b_\n^k~.
\end{equation}
The Green's function then reads
\begin{equation}
\langle J_B^\m(-k) J_B^\n (k)\rangle_R = G_R^{\m\n}(k) =  \frac1{2\ka_5^2\,q^2}\lim_{u\to0} \left[ \sqrt{-g}\,a\,g^{uu}  u^{2\Xi-1}f_\r^{-k\,\m}K^{\r\s}\(\Xi f_\s^{k\,\n} + u f_\s^{\prime k\,\n}\)\right]\,,
\end{equation}
and its imaginary part
\begin{equation}
{\rm Im}[G_R^{\m\n}(k)] =  -\frac\rmi{4\ka_5^2\,q^2}\lim_{u\to0} \left[ \sqrt{-g}\,a\,g^{uu}  u^{2\Xi}K^{\r\s}\( f_\r^{-k\,\m}\,f_\s^{\prime k\,\n}-f_\r^{k\,\m}\,f_\s^{\prime -k\,\n}\)\right]\,,
\label{imgra}
\end{equation}
is essentially \eqref{conserved}, and therefore independent from $u$. Here we are interested in the longitudinal component (\ref{GLL}) of the above correlator. 
The imaginary part of this quantity coincides with the conserved current in \eqref{conserved} when the latter is computed on solutions satisfying the boundary condition $b_\m^k = k_\m$. In position space the boundary condition would read $B_\m(u\to0,x)= \del_\m C(x)$.

Since the right hand side of (\ref{imgra}) is constant we can evaluate the longitudinal correlator at the horizon $u=1$. We obtain
\begin{align}
{\rm Im}[G_R^{LL}(k)] =& \frac{a_h\om\ell}{2\ka_5^2q^2 z_h^3}k_\m k_\n\times {\cal F}^{\m\n}\,, \nn
{\cal F}^{\m\n}\equiv& \left\{\frac{(q\ell)^2H_h+a_hz_h^2|\vec k|^2}{a_h\om^2}  f_{h\,t}^{-k\,\m}f_{h\,t}^{k\,\n} + z_h^2  f_{h\,i}^{-k\,\m}f_{h\,i}^{k\,\n} +z_h^2\frac{k^i}{\om} \( f_{h\,t}^{-k\,\m}f_{h\,i}^{k\,\n} +f_{h\,i}^{-k\,\n}f_{h\,t}^{k\,\m}  \) \right\}	\,,	\nn
\Gamma_{CS}  =&\lim_{|{\vec k}|,\om\to0}\frac{a_h\ell}{\pi\ka_5^2q^2 z_h^4}k_\m k_\n\times {\cal F}^{\m\n} =\frac{H_h \pi^3\ell^3}{\ka_5^2}T^4  |f_{h\,t}^{k\,t}|^2 \,,\label{Gammahere}
\end{align}
where we assumed all the $f$'s are analytic in $k^i$ (before taking the zero-frequency limit). 
All we need to compute the CS diffusion rate is the norm of the coefficient $f_{h\,t}^{k\,t}$ at $|\vec k|=\om=0$. 
However, it turns out that for the temporal component of a massive bulk vector the coefficient $f_{h\,t}^{k\,t}$ always goes to zero with $\om$ (at zero momentum), and therefore formula \eqref{CSdefinition} gives zero.

In fact, the solution to the equations of motion admits the following expansion close to the horizon\footnote{We suppress the super and subscript $t$.} (this is model independent):
\begin{equation}
f^{k} = (1-u)^{-\rmi\frac{w}{4}} \left[ f_{h}^{k} + (1-u) f_{h(1)}^{k}  + \cO(1-u)^2\right]\,.
\end{equation}
On the AdS-Schwarzschild background one then finds 
\begin{equation}
f_{h(1)}^{k}  = \frac{ \(-16 \Xi(\Xi -1) +4 \rmi \Xi  (\Xi +1) w +(4 \Xi -1) w^2 -\rmi w^3 \)H_h}{4 w (w+2 \rmi)}  f_{h}^{k}\,.
\end{equation}
For small $w$ (and $q\neq0$) the above formula gives
\begin{equation}
f_{h(1)}^{k}\sim \rmi\frac{q^2H_h}{2w }f_{h}^{k}\,,
\end{equation}
which explodes unless $f_{h}^{k}\sim w$,\footnote{The expressions for $f_{h(2)}^{k}$ and $f_{h(3)}^{k}$ all suffer from the same issue.} so that from (\ref{Gammahere}) it follows that $\Gamma_{CS}=0$.


\begin{thebibliography}{}

\bibitem{Fukushima:2008xe} 
  K.~Fukushima, D.~E.~Kharzeev and H.~J.~Warringa,
  ``The Chiral Magnetic Effect,''
  Phys.\ Rev.\ D {\bf 78}, 074033 (2008)
  [arXiv:0808.3382 [hep-ph]].

\bibitem{Moore:2010jd} 
  G.~D.~Moore and M.~Tassler,
  ``The Sphaleron Rate in SU(N) Gauge Theory,''
  JHEP {\bf 1102}, 105 (2011)
  [arXiv:1011.1167 [hep-ph]].
\bibitem{Son:2002sd} 
  D.~T.~Son and A.~O.~Starinets,
  ``Minkowski space correlators in AdS / CFT correspondence: Recipe and applications,''
  JHEP {\bf 0209}, 042 (2002)
  [hep-th/0205051].
\bibitem{Policastro:2001yc} 
  G.~Policastro, D.~T.~Son and A.~O.~Starinets,
  ``The Shear viscosity of strongly coupled N=4 supersymmetric Yang-Mills plasma,''
  Phys.\ Rev.\ Lett.\  {\bf 87}, 081601 (2001)
  [hep-th/0104066].
  
\bibitem{Kovtun:2004de} 
  P.~Kovtun, D.~T.~Son and A.~O.~Starinets,
  ``Viscosity in strongly interacting quantum field theories from black hole physics,''
  Phys.\ Rev.\ Lett.\  {\bf 94}, 111601 (2005)
  [hep-th/0405231].
 \bibitem{matranca} D.~Mateos and D.~Trancanelli,
  ``The anisotropic N=4 super Yang-Mills plasma and its instabilities,''
  Phys.\ Rev.\ Lett.\  {\bf 107}, 101601 (2011)
  [arXiv:1105.3472 [hep-th]].
 D.~Mateos and D.~Trancanelli,
  ``Thermodynamics and Instabilities of a Strongly Coupled Anisotropic Plasma,''
  JHEP {\bf 1107}, 054 (2011)
  [arXiv:1106.1637 [hep-th]].
  \bibitem{bu} Y.~Bu,
  ``Chern-Simons diffusion rate in anisotropic plasma at strong coupling,''
  Phys.\ Rev.\ D {\bf 89}, no. 8, 086003 (2014).
\bibitem{Basar:2012gh} 
  G.~Basar and D.~E.~Kharzeev,
  ``The Chern-Simons diffusion rate in strongly coupled N=4 SYM plasma in an external magnetic field,''
  Phys.\ Rev.\ D {\bf 85}, 086012 (2012)
  [arXiv:1202.2161 [hep-th]].
\bibitem{Witten:1998zw} 
  E.~Witten,
  ``Anti-de Sitter space, thermal phase transition, and confinement in gauge theories,''
  Adv.\ Theor.\ Math.\ Phys.\  {\bf 2}, 505 (1998)
  [hep-th/9803131].
\bibitem{SS} 
  T.~Sakai and S.~Sugimoto,
  ``Low energy hadron physics in holographic QCD,''
  Prog.\ Theor.\ Phys.\  {\bf 113}, 843 (2005)
  [hep-th/0412141].
\bibitem{ben} 
  B.~Craps, C.~Hoyos, P.~Surowka and P.~Taels,
  ``Chern-Simons diffusion rate in a holographic Yang-Mills theory,''
  JHEP {\bf 1211}, 109 (2012)
  Erratum: [JHEP {\bf 1302}, 087 (2013)]
  [arXiv:1209.2532 [hep-th]].
\bibitem{Chamseddine:1997nm} 
  A.~H.~Chamseddine and M.~S.~Volkov,
  ``NonAbelian BPS monopoles in N=4 gauged supergravity,''
  Phys.\ Rev.\ Lett.\  {\bf 79}, 3343 (1997)
  [hep-th/9707176].
  \bibitem{mn} 
  J.~M.~Maldacena and C.~Nunez,
  ``Towards the large N limit of pure N=1 superYang-Mills,''
  Phys.\ Rev.\ Lett.\  {\bf 86}, 588 (2001)
  [hep-th/0008001].
\bibitem{Borsanyi:2013bia} 
  S.~Borsanyi, Z.~Fodor, C.~Hoelbling, S.~D.~Katz, S.~Krieg and K.~K.~Szabo,
  ``Full result for the QCD equation of state with 2+1 flavors,''
  Phys.\ Lett.\ B {\bf 730}, 99 (2014)
  [arXiv:1309.5258 [hep-lat]].
\bibitem{Bazavov:2014pvz} 
  A.~Bazavov {\it et al.} [HotQCD Collaboration],
  ``Equation of state in ( 2+1 )-flavor QCD,''
  Phys.\ Rev.\ D {\bf 90}, 094503 (2014)
  [arXiv:1407.6387 [hep-lat]].
\bibitem{Klebanov:2002gr} 
  I.~R.~Klebanov, P.~Ouyang and E.~Witten,
  ``A Gravity dual of the chiral anomaly,''
  Phys.\ Rev.\ D {\bf 65}, 105007 (2002)
  [hep-th/0202056].
\bibitem{Drwenski:2015sha} 
  T.~Drwenski, U.~Gursoy and I.~Iatrakis,
  ``Thermodynamics and CP-odd transport in Holographic QCD with Finite Magnetic Field,''
  JHEP {\bf 1612}, 049 (2016)
  [arXiv:1506.01350 [hep-th]].
\bibitem{Iatrakis:2015fma} 
  I.~Iatrakis, S.~Lin and Y.~Yin,
  ``The anomalous transport of axial charge: topological vs non-topological fluctuations,''
  JHEP {\bf 1509}, 030 (2015)
  [arXiv:1506.01384 [hep-th]].
 \bibitem{kiritsis} 
  U. Gursoy, I.~Iatrakis, E.~Kiritsis, F.~Nitti and A.~O'Bannon,
  ``The Chern-Simons Diffusion Rate in Improved Holographic QCD,''
  JHEP {\bf 1302}, 119 (2013)
  [arXiv:1212.3894 [hep-th]].
\bibitem{Bigazzi:2009bk} 
  F.~Bigazzi, A.~L.~Cotrone, J.~Mas, A.~Paredes, A.~V.~Ramallo and J.~Tarrio,
  ``D3-D7 Quark-Gluon Plasmas,''
  JHEP {\bf 0911}, 117 (2009)
  doi:10.1088/1126-6708/2009/11/117
  [arXiv:0909.2865 [hep-th]].
\bibitem{Faedo:2016cih} 
  A.~F.~Faedo, D.~Mateos, C.~Pantelidou and J.~Tarrio,
  ``Holography with a Landau pole,''
  JHEP {\bf 1702}, 047 (2017)
  doi:10.1007/JHEP02(2017)047
  [arXiv:1611.05808 [hep-th]].
\bibitem{Cotrone:2012um} 
  A.~L.~Cotrone and J.~Tarrio,
  ``Consistent reduction of charged D3-D7 systems,''
  JHEP {\bf 1210}, 164 (2012)
  [arXiv:1207.6703 [hep-th]].
   \bibitem{tranca} V.~Jahnke, A.~S.~Misobuchi and D.~Trancanelli,
  ``Chern-Simons diffusion rate from higher curvature gravity,''
  Phys.\ Rev.\ D {\bf 89}, no. 10, 107901 (2014)
  [arXiv:1403.2681 [hep-th]].
 \bibitem{gkt}S.~S.~Gubser, I.~R.~Klebanov and A.~A.~Tseytlin,
  ``Coupling constant dependence in the thermodynamics of N=4 supersymmetric Yang-Mills theory,''
  Nucl.\ Phys.\ B {\bf 534}, 202 (1998)
  [hep-th/9805156].
\bibitem{Klebanov:1998hh} 
  I.~R.~Klebanov and E.~Witten,
  ``Superconformal field theory on three-branes at a Calabi-Yau singularity,''
  Nucl.\ Phys.\ B {\bf 536}, 199 (1998)
  [hep-th/9807080].
\bibitem{Klebanov:2000hb} 
  I.~R.~Klebanov and M.~J.~Strassler,
  ``Supergravity and a confining gauge theory: Duality cascades and chi SB resolution of naked singularities,''
  JHEP {\bf 0008}, 052 (2000)
  [hep-th/0007191].
\bibitem{Danielsson:1999zt} 
  U.~H.~Danielsson, E.~Keski-Vakkuri and M.~Kruczenski,
  ``Spherically collapsing matter in AdS, holography, and shellons,''
  Nucl.\ Phys.\ B {\bf 563}, 279 (1999)
  [hep-th/9905227].
 \bibitem{Jimenez-Alba:2014iia}
  A.~Jimenez-Alba, K.~Landsteiner and L.~Melgar,
  ``Anomalous magnetoresponse and the Stückelberg axion in holography,''
  Phys.\ Rev.\ D {\bf 90} (2014) 126004
  [arXiv:1407.8162 [hep-th]].
 \bibitem{benincasa} 
  P.~Benincasa and A.~Buchel,
  ``Hydrodynamics of Sakai-Sugimoto model in the quenched approximation,''
  Phys.\ Lett.\ B {\bf 640}, 108 (2006)
  [hep-th/0605076].
 \bibitem{noiwym} 
  F.~Bigazzi, A.~L.~Cotrone, L.~Martucci and L.~A.~Pando Zayas,
  ``Wilson loop, Regge trajectory and hadron masses in a Yang-Mills theory from semiclassical strings,''
  Phys.\ Rev.\ D {\bf 71}, 066002 (2005)
  [hep-th/0409205].
 \bibitem{gtv} 
  S.~S.~Gubser, A.~A.~Tseytlin and M.~S.~Volkov,
  ``NonAbelian 4-d black holes, wrapped five-branes, and their dual descriptions,''
  JHEP {\bf 0109}, 017 (2001)
  [hep-th/0108205].
   \bibitem{divecchia}
  P.~Di Vecchia, A.~Lerda and P.~Merlatti,
  ``N=1 and N=2 superYang-Mills theories from wrapped branes,''
  Nucl.\ Phys.\ B {\bf 646}, 43 (2002)
  [hep-th/0205204].
  \bibitem{bertolini} 
  M.~Bertolini and P.~Merlatti,
  ``A Note on the dual of N = 1 superYang-Mills theory,''
  Phys.\ Lett.\ B {\bf 556}, 80 (2003)
  [hep-th/0211142].
  \bibitem{noi}
  F.~Bigazzi, A.~L.~Cotrone, M.~Petrini and A.~Zaffaroni,
  ``Supergravity duals of supersymmetric four-dimensional gauge theories,''
  Riv.\ Nuovo Cim.\  {\bf 25N12}, 1 (2002)
  [hep-th/0303191].
 \bibitem{cassani} 
  D.~Cassani and A.~F.~Faedo,
  ``A Supersymmetric consistent truncation for conifold solutions,''
  Nucl.\ Phys.\ B {\bf 843}, 455 (2011)
  [arXiv:1008.0883 [hep-th]].
\end{thebibliography}
\end{document}